\documentclass{PoS}

\title{IAXO, next-generation of helioscopes}

\ShortTitle{IAXO}

\author{\speaker{M. Giannotti}\\
        Physical Sciences, Barry University,
        11300 NE 2nd Ave., Miami Shores, FL 33161, USA\\
        E-mail: \email{mgiannotti@mail.barry.edu}}

\author{J. Ruz\\
        Lawrence Livermore National Laboratory, 7000 East Ave., Livermore, CA, 94550, USA\\
        E-mail: \email{ruzarmendari1@llnl.gov}}

\author{J. K. Vogel\\
        Lawrence Livermore National Laboratory, 7000 East Ave., Livermore, CA, 94550, USA\\
        E-mail: \email{vogel9@llnl.gov}}
    
\author{On behalf of the IAXO Collaboration}

\abstract{The International Axion Observatory (IAXO) is a forth generation axion helioscope designed to detect solar axions and axion-like particles (ALPs) with a coupling to the photon $g_{a\gamma}$ down to a few  $10^{-12}$~GeV$^{-1}$, 1.5 orders of magnitude beyond the current best astrophysical and experimental upper bounds. This range includes parameter values invoked in the context of the observed anomalies in light propagation over astronomical distances and to explain the excessive cooling observed in a number of stellar objects. 
Here we review the status of the IAXO project and of its potential to probe the most physically motivated regions of the axion/ALPs parameter space. 
}

\FullConference{38th International Conference on High Energy Physics\\
		3-10 August 2016\\
		Chicago, USA}

\begin{document}

\section{Introduction}
The interest in axions~\cite{Weinberg:1977ma,Wilczek:1977pj}, axion-like-particles (ALPs) and, more generally, in light, weakly interactive particles~\cite{Jaeckel:2010ni} has substantially grown in recent years. 
The enthusiasm can be ascribed to the observation of several anomalies in particle physics, astrophysics and cosmology, whose solutions have invoked this kind of new physics  and to the experimental advancements which are finally allowing the investigation of the most interesting regions of the axion/ALP parameter space.
 
\begin{figure}[t]
	\begin{center}
		\includegraphics[width=15cm]{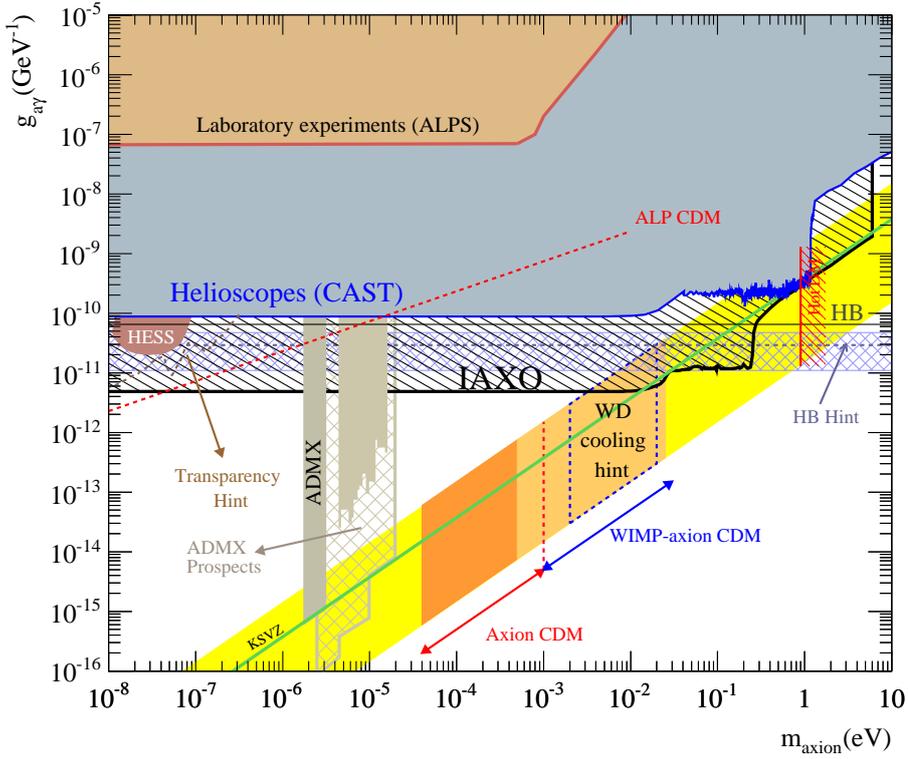}
		\caption{Summary of axion/ALP searches, phenomenological hints and bounds (updated from ref.~\cite{Carosi:2013rla}).}
		\label{fig:sensitivity}
	\end{center}
\end{figure}

A predominant motivation for the axion is the still unresolved strong CP problem~\cite{Peccei:1977hh,Peccei:1977ur}, that is the lack of observation of the expected CP violating effects in the strong interactions. 
The Peccei-Quinn mechanism~\cite{Peccei:1977hh,Peccei:1977ur}, arguably the most appealing solution of this problem, predicts the axion as the pseudo Goldstone boson associated with the spontaneous breaking of a novel Abelian symmetry of the Lagrangian.
Such a particle is expected to couple to standard model fields and, in particular, to photons according to the following Lagrangian term
\begin{equation}\label{Eq:agg}
L=-\frac{g_{a\gamma}}{4} a F\tilde F=g_{a\gamma}\,a \vec E\cdot\vec B \,. 	
\end{equation}
The axion mass is proportional to the photon coupling, $ (m_{a}/1\mbox{ eV})= 0.5\times 10^{10}{\rm GeV}\,\xi \,g_{a\gamma}$ 
where $\xi\simeq 1 $ in many motivated axion models. 
This relation describes a strip (the width given by reasonable values of the model dependent parameters) in the mass-photon coupling parameter space, known as the QCD axion band (yellow band in Fig.~\ref{fig:sensitivity}).
Belonging to this, however, is not a requirement for the solution of the strong CP problem~\cite{Rubakov:1997vp,Berezhiani:2000gh,Gianfagna:2004je}.
Moreover, light pseudoscalar Axion Like Particles (ALPs), weakly coupled to photons as in (\ref{Eq:agg}) but outside the QCD axion band, emerge naturally in various extensions of the Standard Model (see, e.g., \cite{Jaeckel:2010ni,Ringwald:2012hr,Baker:2013zta}). 

Besides particle physics, axions and ALPs may play a fundamental role in astrophysics and cosmology. 
In particular, they are excellent candidates for the cold dark matter in the universe~\cite{Abbott:1982af,Dine:1982ah,Preskill:1982cy,Arias:2012az}
and could explain the anomalous cooling of several stellar systems~\cite{Giannotti:2015kwo}, 
the seeming transparency of the universe to very high energy (TeV) gamma rays in the galactic and extragalactic medium~\cite{Horns:2012fx}, some anomalous redshift-dependence of AGN gamma-ray spectra~\cite{Galanti:2015rda} and, perhaps,  
anomalous x-ray observations of the active Sun~\cite{Rusov:2015sqa}.  

All these observations, which span particle physics, astrophysics and cosmology, have motivated several new experimental efforts to study the most relevant axion/ALP parameter space. 
Among those, the International Axion Observatory (IAXO)~\cite{JCAP 06 (2011) 013,2014 JINST 9 T05002,Vogel:2015yka} stands out as the only technology which can allow the study of a wide mass range in the ALP parameter space all the way into the parameter region predicted by well motivated QCD axion models, and down to coupling constants small enough to include the parameters predicted by the various astrophysical anomalies. 

In this contribution we update the status of the IAXO project and discuss its sensitivity potential with respect to other experiments and astrophysical observations.

\section{Axions and Axion Like Particles: Astrophysics and Experimental Searches}
\label{sec:Axions_ALPs}
%
Stars have provided a unique and extremely successful laboratory to study axions and ALPs, leading throughout the years to strong constraints on their couplings to photons~\cite{Friedland:2012hj,Carosi:2013rla,Ayala:2014pea},\footnote{Even stronger bounds have been derived for very low masses~\cite{Payez:2014xsa,TheFermi-LAT:2016zue,Berg:2016ese}.
A recent update of the method in ref.~\cite{Payez:2014xsa} to probe the axion-photon coupling in the case of a future galactic supernova is given in~\cite{Meyer:2016wrm}.} 
 electrons~\cite{Viaux:2013lha}, and nucleons~\cite{Sedrakian:2015krq,Fischer:2016cyd}.
More intriguingly, several independent observations of the cooling of different stellar systems seem to agree with the need for a novel cooling mechanism, such as the one provided by axions or ALPs, in addition to the standard astrophysical processes (see M.Giannotti's contribution to these proceedings).
Though this is still only a speculative hypothesis, it is remarkable that axions or ALPs are unique, among the several weakly interactive particle candidates, in their ability to explain all the independent observations~\cite{Giannotti:2015kwo}.

Terrestrial experiments aimed at detecting axions and ALPs rely, in the large majority of cases, on the electromagnetic coupling (\ref{Eq:agg}) which allows for ALPs to transform into photons in a strong magnetic field~\cite{Sikivie:1983ip}.
The main experimental strategies which make use of this idea may be classified into three categories:
\textit{i)} \textit{Light shining through wall} experiments, such as the planned laser experiment ALPS II~\cite{Bahre:2013ywa}, particularly efficient at low masses.
\textit{ii) Haloscopes}, such as ADMX~\cite{Hoskins:2016svf} and ADMX-HF~\cite{Brubaker:2016ktl}, which look for the conversion of axions from the local dark matter into microwave photons. Haloscopes are very sensitive, though only in narrow mass ranges. 
Their sensitivity depends on the unknown axion fraction of the local DM.
\textit{iii) Helioscopes}, which look for x-ray photons from the conversion of ALPs produced in the core of the sun.
Among these techniques the axion heliscope has the clear advantage to allowing the exploration of a wide mass range.


\section{A forth generation axion helioscope}
\label{sec:IAXO}
Axion helioscopes are searching for axions produced in the solar core employing a strong laboratory magnetic field to reconvert these axions into x-ray photons. 
The technology has evolved over the last two decades and culminated in the CERN Axion Solar Telescope (CAST), a third generation helioscope, which pushed the bounds on the photon coupling below any other terrestrial experiment in a large mass range, reaching for the first time the astrophysical bounds and successfully exploring a section of the QCD axion band.

All three generations of axion helioscopes~\cite{Brookhaven,TokyoHelioscope, CAST} have used existing magnets that were designed and built for other purposes and delivered sensitivity improvements of about a factor of $5-7$ from one generation to the next. 
While considerably saving on the cost, this strategy has severely limited the helioscope physics potential.
For IAXO~\cite{JCAP 06 (2011) 013}, a fourth generation helioscope, the magnet is specifically designed to maximize the sensitivity while keeping low cost and minimal technical
challenges.

The envisioned magnet for IAXO plays an important role in significantly improving the sensitivity of the experiment to the axion-photon coupling as the figure of merit (FOM) study in Ref.~\cite{JCAP 06 (2011) 013} shows.
Given that the currently most suitable, cost-saving, available large scale magnets are based on NbTi superconductor technology providing maximal magnetic fields of up to $6$~T, the magnet parameter to scale up for IAXO is its cross-sectional area. According to first optimization studies~\cite{JCAP 06 (2011) 013} an ATLAS-inspired toroidal geometry provides the best approach to reach IAXO's sensitivity goal. The preliminary design foresees up to eight coils of $21$~m length and $1$~m width yielding an overall magnet system which is $5.2$~m in diameter and $25$~m long. With a stored energy of $500$~MJ (operational current $12.3$~kA) it will be able to provide magnetic fields up to $5.4$~T. Further details can be found in Ref.~\cite{2014 JINST 9 T05002}.
%

Each of the eight IAXO magnet bores will be covered by an x-ray telescope that focuses the x-rays from a putative axion signal into a $\sim 0.2$~cm$^2$ spot which is imaged on an extremely low-background detector positioned in the focal plane of each optic. This will allow an additional increase in sensitivity by enabling the use of small-area detectors which in turn will result in a lower background level, essential for a rare-event search of the IAXO-type. The favored technological approach for the telescopes are segmented, slumped glass optics used in combination with multilayer coatings that enhance the reflectivity of the telescopes. Not only is this technology~\cite{NuSTAR} mature and cost-effective, it also is able to fulfill the IAXO requirements in terms of throughput and spatial resolution. While the radius of the telescopes will vary from $50$~mm to $300$~mm fully covering the magnet aperture, the optimal focal length is going to be about $5$~m.  
%

The IAXO collaboration selected small-area Time Projection Chambers (TPCs) with a pixelated Micromegas readout as the baseline technology for the experiment. These kind of detectors, which rely on a manufacturing technique referred to as microbulk technology, have already been extensively used in the CAST experiment and are able to achieve very low levels of background. At CAST, background rates below $10^{-6}$~counts keV$^{-1}$ cm$^{-2}$ s$^{-1}$ (ckcs) have been demonstrated while underground laboratory tests indicate that levels of  $10^{-7}$~ckcs and below are feasible. Such levels (as low as $10^{-8}$~ckcs) are the ultimate goal for IAXO, since this would provide a quasi-zero background situation when taking into account the proposed exposure times for IAXO.
%

The team recently built, implemented and characterized a pathfinder system for IAXO consisting of a multilayer-coated x-ray telescope coupled to one of the above mentioned ultra-low background Micromegas detectors~\cite{Aznar:2015iia}. The system was installed at CAST and acquired data in axion-sensitive conditions during 2014/15 (publication in preparation). A full characterization of the optic including measurements of the spatial resolution and the throughput has recently been completed successfully at the PANTER x-ray test facility of the Max-Planck Institute in Garching, Germany, and the overall results show that the approach of using segmented glass optics similar to those currently flying onboard NASA's NuSTAR mission are able to meet IAXO requirements. Full results of the calibration campaign will be published shortly.

\section{Discussion}

%
Axion helioscopes stand out among other axion/ALP detection techniques as the most mature, technologically viable, and scalable experiments. 
Though they do not reach the sensitivity of axion haloscopes, helioscopes are independent of axion mass up to a relatively large value, with the
addition of a buffer gas expanding the sensitivity to even higher masses, while haloscopes can probe only very narrow mass ranges.
Additionally, in contrast to haloscopes, they do not rely on assumptions of axions being the dominant component of dark matter. 

IAXO will build on the experience of the current world-leading helioscope experiment, CAST, to improve sensitivity to axion signals by over 5 orders of magnitude, probing the ALP-photon coupling down to a few $ 10^{-12} $ GeV$ ^{-1} $, for masses up to 0.25 eV (see, Fig.~\ref{fig:sensitivity}). 
%
%
For masses down to a few meV, it will be sensitive to a broad range of QCD axion models that could constitute
all or part of the dark matter in the universe. 
ALPs, WISPs, or non-standard cosmological frameworks
broaden the parameter space to include additional dark matter candidates also within reach of IAXO. 

Finally, IAXO has the potential to provide a definitive test of the ALP invoked to account for anomalous light propagation over large distances and to probe some of the most relevant parameter space hinted by the stellar cooling anomalies.
A coupling to electrons of the size expected in non-hadronic axion models would largely enhance the solar axion/ALP production rate~\cite{Redondo:2013wwa} 
%
%
making IAXO also (indirectly) sensitive to the axion-electron coupling in the solar core and allowing the detection of axions or ALPs potentially responsible for the excessive cooling of white dwarfs and red giants.

After the negative results of the LHC searches for a massive dark matter candidate, perhaps we should look elsewhere, to the low energy frontier. 
IAXO offers a unique possibility to explore this option.  

\acknowledgments
We are grateful to the organizers for the very interesting conference.
We also wish to thank our colleagues I. Irastorza, B. Lakic, A. Lindner and A. Ringwald for their feedback and for many stimulating comments.
Part of this work was performed under the auspices of the US Department of Energy by Lawrence Livermore National Laboratory under Contract DE-AC52-07NA27344 and LDRD 17-ERD-030.

\end{document}